\documentclass{ws-ijmpcs}

\begin{document}

\markboth{Z.T. Liang}
{Three dimensional imaging of the nucleon  --- TMD ... }

%
\catchline{}{}{}{}{}
%

\title{Three dimensional imaging of the nucleon\\
--- TMD (theory and phenomenology)}

\author{Zuo-tang Liang}

\address{School of physics and key laboratory of particle physics \& particle irradiation (MOE), \\
Shandong University, Jinan, Shandong 250100, CHINA}

\maketitle

\begin{history}
\received{Day Month Year}
\revised{Day Month Year}
\published{Day Month Year}
\end{history}

\begin{abstract}
This is intend to provide an overview of the theory and phenomenology parts of the TMD 
(Transverse Momentum Dependent parton distribution and fragmentation functions) studies.
By comparing with the theoretical framework that we have for the inclusive deep inelastic lepton-nucleon 
scattering and the one-dimensional imaging of the nucleon, 
I try to outline what we need to do in order to construct a comprehensive 
theoretical framework for semi-inclusive reactions and the three dimensional imaging of the nucleon. 
After that, I try to give an overview of what we have already achieved and make an outlook for the future. 

\end{abstract}

\ccode{PACS numbers: 12.38.-t, 12.38.Bx, 12.39.St, 13.60.-r, 13.66.Bc, 13.87.Fh, 13.88.+e, 13.40.-f, 13.85.Ni}

\section{Introduction}	

With the deep going of the study of the nucleon structure, three dimensional imaging has become the frontier and a hot topic in recent years.   
It is commonly recognized that the three dimensional imaging contains much more abundant physics  on the nucleon structure and the properties of QCD. 
The study was initially triggered by the experimental finding of striking single-spin asymmetries (SSA) in inclusive hadron production in hadron-hadron collisions. 
Gradually it grows into a field aiming at a comprehensive picture of nucleon structure including spin and transverse momentum dependences.  

The one dimensional imaging of the nucleon is provided by the parton distribution functions such as the number densities, $q(x)$, 
the helicity distributions, $\Delta q(x)$, and the transversities, $\delta q(x)$, for quarks of different flavors.  
In the three dimensional case, i.e. where the transverse momentum is also considered, not only the direct extensions of them to include transverse momenta 
are involved, but also many other correlation functions such as the Sivers function, the Boer-Mulders function, the pretzelocity etc exist. 
Moreover, higher twist effects become also important and need to be considered consistently. 

The study on the three dimensional imaging of the nucleon is in a rapid developing phase and it is not so easy to present 
a comprehensive overview of all different aspects of studies. 
Here, I choose to do the job in the following way: 
First, I will try to make a brief review of what we did in one dimensional case with inclusive DIS. 
In this way, I hope that I can show you the main line of what we need to do in three dimensional case. 
Then I will try to summarize what we have already achieved along this line and what we need to do next. 
For the sake of space, I will mainly concentrate on the discussions but keep as less equations as possible. 
An extended version is prepared and will be published in a special issue of Frontier of Physics.

 \section{Inclusive DIS \& the One Dimensional Imagining of the Nucleon}
 
Our studies on the structure of a fast moving nucleon started with the inclusive deep inelastic scattering  (DIS) 
e.g. $e^-+N\to e^-+X$. We recall that, under one photon exchange, the differential cross section is given 
by the Lorentz contraction of the known leptonic tensor and the hadronic tensor $W_{\mu\nu}(q,p,S)$  
(where $p$ and $S$ are the 4-momentum and polarization vector of the nucleon).
Information of the structure of the nucleon is contained in the hadronic tensor $W_{\mu\nu}(q,p,S)$. 


The theoretical framework for inclusive DIS has been constructed in the following steps. 
First, we studied the kinematics and obtained the general form of the hadronic tensor by applying the 
basic constraints from the general symmetry requirements such as Lorentz covariance, gauge invariance, parity conservation and Hermiticity. 
We found out that the hadronic tensor is determined by four independent structure functions $F_1$, $F_2$, $g_1$ and $g_2$, 
where the first two describe the unpolarized case and the latter two are needed for polarized cases.

Our knowledge of one dimensional imaging of the nucleon starts with the ``intuitive parton model" that is very nicely formulated e.g. in [\refcite{Fey72}]. 
Here, it was argued that, in a fast moving frame, because of time dilation, quantum fluctuation such as vacuum polarizations, 
can exist quite long. In the infinite momentum frame, such fluctuations exist for ever. In this case, a fast moving nucleon can be 
viewed as a beam of free ``partons". The probability of the scattering of an electron with the nucleon is equal to 
that of the scattering with a parton convoluted with the number density of the parton in the nucleon, i.e., 
\begin{align}
|{\cal M}(eN\to eX)|^2=\sum_q\int dx f_q(x) |\hat{\cal M}(eq\to eq)|^2, \label{eq:intuitive}
\end{align}
where $f_q(x)$ is the number density of $q$ 
in the nucleon. 
In this way, we obtained the famous results, 
$F_2(x,Q^2)=2xF_1(x,Q^2)=\sum_q e_q^2xf_q(x),$ 
$g_1(x,Q^2)=\sum_q e_q^2x\Delta f_q(x)$ and so on. 
Here, I would like to point out that, with this intuitive parton model, we are doing nothing else but the impulse approximation that we often used 
in describing a collision process where we do the following approximations,
\begin{itemize}
\item during the interaction of the electron with the parton, interactions between the partons are neglected;
\item the scatterings of the electron with different partons are added incoherently;
\item the electron interacts only with one single parton.
\end{itemize}

\begin{figure}[pb]
\vspace*{-20pt}
\centerline{\includegraphics[width=7.7cm]{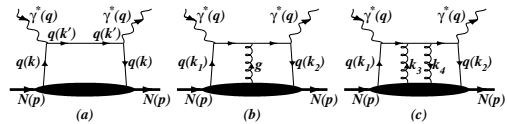}}
\vspace*{-12pt}
\caption{Examples of the Feynman diagram series with multiple gluon scattering considered 
  for $\gamma^*+N\to q+X$ with (a) $j=0$, (b) $j=1$ and (c) $j=2$ gluons exchanged.}
\end{figure}

Although the intuitive model is elegant and practical, we are not satisfied since it is not easy to control the accuracy.  
The proper quantum field theoretical (QFT) formulation is given by starting with Feynman digram Fig.~1(a)
where we obtain, 
\begin{align}
&W_{\mu\nu}^{(0)}(q,p,S)=\frac{1}{2\pi}
\int\frac{d^4k}{(2\pi)^4} 
{\rm Tr}[\hat H_{\mu\nu}^{(0)}(k,q)\hat \phi^{(0)}(k,p,S)],\label{eq:Wincl0}
\end{align}
where $k$ is the 4-momentum of the parton,  
$\hat H_{\mu\nu}^{(0)}(q,k)=
\gamma_\mu(\slash{\hspace{-5pt}k}+\slash{\hspace{-5pt}q})\gamma_\nu
(2\pi)\delta_+((k-q)^2) $ 
is a calculable hard part and the matrix element 
\begin{align}
& \hat\phi^{(0)}(k,p,S) =  
\int d^4ze^{ikz} \langle p,S|\bar\psi(0)\psi(z)|p,S\rangle, \label{eq:phi0}
\end{align}
is known as the quark-quark correlator describing the structure of the nucleon.  
By taking the collinear approximation, i.e. taking $k\approx xp$, and neglecting 
the power suppressed contributions, we obtain exactly the same result as that obtained using Eq. (\ref{eq:intuitive}) based on the intuitive parton model.
[Here, we use light-cone coordinate $k^\mu=(k^+,k^-,\vec k_\perp)$ and take
$\bar n=(1,0,\vec 0_\perp)$, $n=(0,1,\vec 0_\perp)$, $n_{\perp}=(0,0,\vec n_\perp)$. 
Also, we choose the coordinate system such that $p=p^+\bar n$.]  
At the same time, we obtain also a QFT operator expression of $f_q(x)$,  
\begin{align}
f_q(x)=\int \frac{dz^-}{2\pi}e^{ixp^+z^-}\langle p|\bar\psi(0)\frac{\gamma^+}{2}\psi(z)|p\rangle,
\end{align}
which is indeed the number density of parton in the nucleon. 
However, from this expression, we see immediately that it is not (local) gauge invariant! 
To get the gauge invariant formulation, we need to take into account the multiple gluon scattering given by the diagram series shown in Fig.1(a-c).
The contribution from each diagram is expressed as a trace of a hard part and a matrix element. E.g., for Fig. 1(b), 
\begin{align}
&W_{\mu\nu}^{(1,L)}(q,p,S)=\frac{1}{2\pi}
\int\frac{d^4k_1}{(2\pi)^4} \frac{d^4k_2}{(2\pi)^4} 
{\rm Tr}[\hat H_{\mu\nu}^{(1,L)}(k_1,k_2,q)\hat \phi^{(1)}_\rho(k_1,k_2,p,S)],\label{eq:Wincl1L}\\
&\hat \phi^{(1)}_\rho(k_1,k_2,p,S)=\int d^4z d^4y e^{ik_1z+(k_2-k_1)y} \langle p,S|\bar\psi(0)A_\rho(y)\psi(z)|p,S\rangle, \label{eq:phi0}
\end{align}  
where $L$ in the superscript denotes left cut. 
The matrix element is now a quark-$j$-gluon(s)-quark correlator, where $j$ is the number of gluons.
We also see that none of such quark-$j$-gluon(s)-quark correlators is gauge invariant.

To get the gauge invariant form, we need to apply the collinear expansion 
as proposed in Refs. [\refcite{Ellis:1982wd}-\refcite{Qiu:1990xxa}], which is carried out in the following four steps.

(1) Make Taylor expansions of all hard parts at $k_i = x_ip$, e.g., 
\begin{align}
&\hat H_{\mu\nu}^{(0)}(k,q) =  \hat H_{\mu\nu}^{(0)} (x) + \frac{\partial \hat H_{\mu\nu}^{(0)}(x)}{\partial k_\rho} \omega_\rho^{\ \rho'} k_{\rho'} + \cdots, 
\end{align}
where $\omega_\rho^{\ \rho'}$ is a projection operator defined by $\omega_\rho^{\ \rho'} \equiv g_\rho^{\ \rho'} -\bar n_\rho n^{\rho'}$.

(2) Decompose the gluon field into 
$A_\rho(y) = A^+(y) \bar n_\rho + \omega_\rho^{\ \rho'} A_{\rho'}(y)$.

(3) Apply the Ward identities such as,
\begin{align}
& \frac{\partial \hat H^{(0)}_{\mu\nu}(x)}{\partial k_\rho}=  - \hat H^{(1)\rho}_{\mu\nu} (x,x), \ \ \ \ \ \ 
p_\rho \hat H^{(1,L)\rho}_{\mu\nu}(x_1,x_2) = \frac{H^{(0)}_{\mu\nu} (x_1)}{x_2-x_1-i\epsilon}. 
\end{align}

(4) Add all terms with the same hard part together and we obtain 
$W_{\mu\nu}(q,p,S) = \sum_{j,c}\tilde{W}^{(j,c)}_{\mu\nu}(q,p,S)$ ($j=0,1,2,\cdots$ and $c$ is the different cut) where e.g. for $j=0$,  
\begin{align}
&\tilde W_{\mu\nu}^{(0)}(q,p,S)=\frac{1}{2\pi} \int \frac{d^4k}{(2\pi)^4}
{\rm Tr}\big[\hat H_{\mu\nu}^{(0)}(x)\ \hat \Phi^{(0)}(k,p,S)\big],\label{eq:tWincl0}\\
&\hat\Phi^{(0)}(k,p,S)=\int {d^4y} e^{iky} \langle p,S|\bar{\psi}(0){\cal{L}}(0;y)\psi(y)|p,S\rangle,  \label{eq:Phi0def}
\end{align}
where 
${\cal{L}}(0;y)=\mathcal{L}^\dag(\infty; 0) \mathcal{L}(\infty; y)$, and  
${\cal{L}}(\infty; y)= P e^{- i g \int_{y^-}^\infty d \xi^{-} A^+ ( \xi^-, \vec{y}_{\perp})}$ 
($P$ stands for path integral), 
is the well-known gauge link obtained in the collinear expansion.

In this way, we have constructed the theoretical framework for calculating the contributions 
at the leading order (LO) in pQCD but leading as well as higher twist contributions in a systematical way. 
The results are given in terms of the gauge invariant parton distribution and correlation functions (PDFs).
We also see that the PDFs involved here are all scale independent. 
This is because we have till now considered only the LO pQCD contributions, i.e. the tree diagrams. 

Because the hard parts in $\tilde W_{\mu\nu}^{(j)}$'s such as that given by Eq.~(\ref{eq:tWincl0})  
are only functions of the longitudinal component $x$ but independent of other components of the parton momentum $k$, 
we can simplify them. E.g., for $j=0$, it reduces to,  
\begin{align}
&\tilde W_{\mu\nu}^{(0)}(q,p,S)=\frac{1}{2\pi} \int {p^+dx}
{\rm Tr}\big[\hat H_{\mu\nu}^{(0)}(x)\ \hat \Phi^{(0)}(x,p,S)\big], \label{eq:tWincl0x}\\
&\hat\Phi^{(0)}(x,p,S)
=\int \frac{dy^-}{2\pi} e^{ixp^+y^-} \langle p,S|\bar{\psi}(0){\cal{L}}(0;y^-)\psi(y^-)|p,S\rangle.  \label{eq:Phi0x}
\end{align}
We see clearly that only one dimensional imaging of the nucleon is relevant in inclusive DIS. 
We also see that the PDFs are defined in terms of QFT operators via the quark-quark correlator $\hat\Phi^{(0)}(x,p,S)$ by expending it 
in terms of $\gamma$-matrices and the corresponding basic Lorentz covariants. 


To go to higher order of pQCD, we take the loop diagrams that describe gluon radiations and so on into account. 
After proper handling of these contributions, we obtain the factorized form\cite{Collins:1989gx} 
where the PDFs acquire the scale $Q$-dependence govern by QCD evolution equations.
In practice, PDFs are parameterized and are given in the PDF library (PDFlib).

In summary, for studying one dimensional imaging of the nucleon with inclusive DIS, we took the following steps,
\begin{itemize}\vspace*{-6pt}
\item General symmetry analysis leads to the general form of the hadronic tensor in terms of four independent structure functions.
\item Parton model without QCD interaction leads to LO in pQCD and leading twist results in terms of 
$Q$-independent PDFs without (local) gauge invariance.
\item  Parton model with QCD multiple gluon scattering after collinear expansion leads to LO in pQCD, leading and 
higher twist contributions in terms of $Q$-independent but gauge invariant PDFs.  
\item Parton model with QCD multiple gluon scattering and ``loop diagram contributions" after collinear approximation, regularization and renormalization 
leads to leading and higher order pQCD, leading twist contributions 
in factorized forms as functions of $Q$-evolved and gauge invariant PDFs.
\end{itemize}

In the following, I will follow this four steps and summarize what we achieved in the three dimensional case. 
Before that, I would like to recall the following two of historical developments that may be helpful for to us 
to construct the theoretical framework for the TMD case. 

First, as mentioned, the study of three dimensional imaging of the nucleon was triggered by 
the experimental observation of SSA in the inclusive hadron-hadron collision. 
It was known that pQCD leads to negligibly small asymmetry for the hard part 
but the observed asymmetry can be as large as 40\%.  
The hunting for such large asymmetries last for decades with following milestones:
\begin{itemize}\vspace*{-6pt}
\item In 1991, Sivers introduced\cite{Sivers:1989cc} the asymmetric quark distribution in a transversely polarized nucleon 
that is now known as the Sivers function. 
\item In 1993, Boros, Liang and Meng proposed\cite{Boros:1993ps} a phenomenological model 
that provides an intuitive physical picture showing that 
the asymmetry arises from the orbital angular momenta of quarks and what they called ``surface effect'' caused by the initial or final state interactions. 
\item In 1993, Collins published\cite{Collins:1992kk} his proof that Sivers function has to vanish. 
\item In 2002, Brodsky, Hwang and Schmidt calculated\cite{Brodsky:2002cx} SSA for SIDIS using an explicit example 
where they took the orbital angular momentum of quark and the multiple gluon scattering into account.
\item Collins pointed out\cite{Collins:2002kn} that the multiple gluon scattering is contained in the gauge link 
and that the conclusion of his proof in 1993 was incorrect because he forgot the gauge link;  
Belitsky, Ji and Yuan resolved\cite{Ji:2002aa,Belitsky:2002sm} the problem of defining the gauge link 
for a TMD parton density in light-cone gauge where the gauge potential does not vanish asymptotically.
\vspace*{-6pt}
\end{itemize}

Another historical development concerns the azimuthal asymmetry study in SIDIS. 
It was shown by Georgi and Politzer in 1977 that\cite{Georgi:1977tv} final state gluon radiations 
lead to azimuthal asymmetries and could be used as a ``clean test to pQCD". 
However, soon after, in 1978, it was shown by Cahn that\cite{Cahn:1978se} similar asymmetries can also 
be obtained if one includes intrinsic transverse momenta of partons. 
The latter, now named as Cahn effect, though power suppressed i.e. higher twist, can be quite significant 
and can not be neglected since the values of the asymmetries themselves are usually not very large. 

The lessens that we learned from these histories are in particular the following two points, i.e., when studying TMDs, 
\begin{itemize}\vspace*{-6pt}
\item it is important to take the gauge link into account;
\item higher twist effects can be important.
\vspace*{-6pt}
\end{itemize}

Both of them demand that, to describe SIDIS in terms of TMDs, we need the proper QFT formulation rather than the intuitive parton model.

\section{TMDs Defined via Quark-Quark Correlator}

The TMD PDFs of quarks are defined via the quark-quark correlator $\Phi^{(0)}$ 
given by Eq. (\ref{eq:Phi0def}) (after integration over $k^-$). 
A systematical study has been given in Ref.[\refcite{Goeke:2005hb}] and a very comprehensive treatment 
can also be found in Ref.[\refcite{Mulders}].
Here, we first expand it in terms of $\gamma$-matrices and obtain 
a scalar, a pseudo scalar, a vector, an axial-vector and a tensor part. 
We then analyze the Lorentz structure of each part by expressing it in terms of possible ``basic Lorentz covariants" and scalar functions. 
These scalar functions are known as TMD PDFs. 
There are totally 32 such TMD PDFs. 
Among them, 8 contribute at leading twist and they all have clear probability interpretations 
such as the number density, the helicity distribution,  the transversity, 
the Sivers function, the Boer-Mulders function etc; 
16 contribute at twist 3 and the other 8 contribute at twist 4. 
We emphasize that they are all scalar functions of $x$ and $k_\perp$, i.e., depending on $x$ and $k_\perp^2$. 
If we integrate over $d^2k_\perp$, terms where the basic Lorentz covariants are space odd vanish. 
At the leading twist, only 3 of 8 survive, i.e. the number density, the helicity distribution and the transversity. 

Higher twist TMD PDFs are also defined via quark-$j$-gluon-quark correlators. 
Many of them are however not independent since they are related to those defined 
via the quark-quark correlator through the QCD equation of motion. 
It is interesting to see that\cite{Song:2013sja}, although not generally proved, 
all the twist 3 TMD PDFs that are defined via quark-gluon-quark correlator 
and are involved in SIDIS 
are replaced by those defined via quark-quark correlator.

I also want to emphasize that fragmentation is just conjugate to parton distribution. 
We have one to one correspondence between TMD PDFs and TMD FFs. 

\section{Accessing the TMDs in High Energy Reactions}

The TMDs can be studied in semi-inclusive high energy reactions such as 
SIDIS $e^-+N\to e^-+h+X$, 
semi-inclusive Drell-Yan $h+h\to l^++l^-+X$, and 
semi-inclusive hadron production in $e^+e^-$-annihilation $e^++e^-\to h_1+h_2+X$. 
With SIDIS, we study TMD PDFs and TMD FFs, 
while with Drell-Yan and $e^+e^-$ annihilation, we study TMD PDFs and TMD FFs separately. 
We now follow the same steps as those for inclusive DIS and briefly summarize what   
we already have in constructing the corresponding theoretical framework. 

(I) The general forms of hadronic tensors: For all three classes of processes,  
the general forms of hadronic tensors have been studied and obtained. 
For SIDIS, it has been discussed in Refs.[\refcite{Gourdin:1973qx}-\refcite{Bacchetta:2006tn}] 
and it has been shown that one need 18 independent structure functions for spinless $h$. 
For Drell-Yan, a comprehensive study was made in Ref. [\refcite{Arnold:2008kf}] 
and the number of independent structure functions is 48 for hadrons with spin 1/2. 
For $e^+e^-$-annihilation,  the study was presented in Ref.[\refcite{Pitonyak:2013dsu}] and one needs 72 for spin-1/2 $h_1$ and $h_2$. 
 
(II) LO in pQCD and leading twist parton model results:    
These are the simplest parton model results and can be obtained easily. E.g., for SIDIS, 
the result can be obtained from those given e.g. in Ref.[\refcite{Bacchetta:2006tn}] by neglecting all the power suppressed contributions. 
I emphasize that the result obtained this way is a complete parton model result at LO in pQCD and leading twist. 
It can be used to extract the TMDs at this order. 
Any attempt to go beyond LO in pQCD or to consider higher twists needs to go beyond this expression. 

(III) LO in pQCD, leading and higher twist parton model results: 
For the semi-inclusive processes where only one hadron is involved, either in the initial or the final state,  
it has been shown\cite{Liang:2006wp}$^-$\cite{Wei:2014pma} 
that the collinear expansion can be applied. 
Such processes include: semi-inclusive DIS $e^-+N\to e^-+q(jet)+X$, and $e^+e^-$-annihilation $e^++e^-\to h+\bar q(jet)+X$. 
By applying the collinear expansion, we have constructed the theoretical frameworks for these processes 
with which leading as well as higher twist contributions can be calculated in a systematical way to LO in pQCD. 
The complete results up to twist-3 have been obtained in Refs.[\refcite{Song:2013sja}-\refcite{Wei:2014pma}].
For unpolarized $e^-+N\to e^-+q(jet)+X$, the results up to twist 4 have also been obtained\cite{Song:2010pf}. 
These results 
can be used as the basis for measuring these TMDs via the corresponding process at the LO in pQCD. 
I call in particular the attention to the results\cite{Wei:2014pma} for $e^++e^-\to h+\bar q(jet)+X$  for $h$ with different spins. 
Those for spin-1 hadrons involve tensor polarization that is much less explored till now. 
See also talks by Y.K. Song and S.Y. Wei for more details\cite{YKSong,SYWei}.

However, for the above-mentioned three kinds of semi-inclusive processes, there are always two hadrons involved. 
Collinear expansion has not been proved how to apply for such processes. 
It is unclear how one can calculate leading and higher twist contributions in a systematical way.
Nevertheless, twist 3 calculations that have been carried out for these processes\cite{Mulders:1995dh,Boer:1997mf,Lu:2011th}, 
practically in the following steps: 
(i) draw Feynman diagrams with multiple gluon scattering to the order of one gluon exchange, 
(ii) insert the gauge link in the correlator wherever needed to make it gauge invariant, 
(iii) carry out calculations to the order $1/Q$. 
Although not proved, it is interesting to see that the results obtained this way reduce exactly to those obtained 
in the corresponding simplified cases where collinear expansion is applied if we take the corresponding fragmentation functions as $\delta$-functions.

(IV) TMD factorization and evolution: 
TMD factorization theorem has been established at the leading twist for 
semi-inclusive processes\cite{Collins:1981uk}$^-$\cite{Ji:2004wu}. 
TMD evolution theory is also developing 
very fast\cite{Henneman:2001ev}$^-$\cite{Echevarria:2014rua}. 
See the overview talk by Daniel Boer\cite{Boer}.
  
\vspace*{-3pt}
\section{TMD Parameterizations}

Experiments have been carried out for all three kinds of semi-inclusive reactions. 
The results are summarized in particular in the plenary talks in this conference
by Marcin Stolarski, Renee Fatemi, and Armine Rostomyan. 
Here, I will just try to sort out the TMD parameterizations that we already have. 

The first part concerns what people called ``the first phase parameterizations", i.e. TMD parameterizations without QCD evolutions. 

(1) Transverse momentum dependence: 
This is usually taken as\cite{Anselmino:2005nn}$^-$\cite{Anselmino:2013lza} 
a Gaussian in a factorized form independent of the longitudinal variable $z$ or $x$. 
The width has been fitted, the form and flavor dependence etc have been tested. 
Roughly speaking, this is a quite satisfactory fit. 
However, it has also been pointed out, e.g. in [\refcite{Signori:2014kda}] for the TMD FF, 
that the Gaussian form seems to depend on the flavor and even on $z$,  
which means that it is only a zeroth order approximation. 

(2) Sivers function: It is usually parameterized\cite{Efremov:2004tp}$^-$\cite{Bacchetta:2011gx} 
in the form of the number density $f_q(x,k_\perp)$ multiplied by a $x$-dependent factor ${\cal N}_q(x)$ and a $k_\perp$-dependent factor $h(k_\perp)$, 
and ${\cal N}_q(x)\sim x^{\alpha}(1-x)^{\beta}$ while $h(k_\perp)$ is taken as a Gaussian. 
There exist already different sets such as 
the Bochum, the Torino and the Vogelsang-Yuan fits. 
One thing seems to be clear that the Sivers function is nonzero for proton and it has different signs for $u$- and $d$-quark.

(3) Transversity and Collins function: 
A simultaneous extraction of them from SIDIS data
have been carried out by the Torino group\cite{Anselmino:2007fs,Anselmino:2013vqa}.
A similar form as that for the Sivers function has been taken 
and it has been obtained that also the Collins function is nonzero and has different signs e.g. for $u\to \pi^+$ or $d\to \pi^+$.

(4) Boer-Mulders function: 
Clear signature for non-zero Boer-Mulders function has been obtained from SIDIS data 
on $\langle\cos2\phi\rangle$ asymmetry\cite{Zhang:2008nu}$^-$\cite{Lu:2009ip}. 
The form was taken again similar to the Sivers function. 
However, I would like to point out that the $\langle\cos2\phi\rangle$ asymmetry receives twist-4 contributions 
due to the Cahn effect\cite{Cahn:1978se}. 
A proper treatment of such twist-4 effect involves twist-4 TMDs as discussed in Ref.~[\refcite{Song:2010pf}]. 
Because of the multiple gluon scattering shown in Fig. 1, the twist-4 effects could be very much different 
from that given in [\refcite{Cahn:1978se}]. 
 
Attempts to parameterize other TMDs such as pretzelocity $h^{\perp}_{1T}$ have also been made\cite{Zhu:2011ir}. 
Although there is no enough data to give high accuracy constraints, the qualitative features obtained are also interesting.

The second part concerns the TMD evolution. 
As mentioned earlier, this is a topic that develops very fast recently. 
A partial list of recent dedicated publications is Refs. [\refcite{Henneman:2001ev}-\refcite{Echevarria:2014rua}]. 
QCD evolution equations have been constructed. 
The numerical results obtained show clearly that TMD evolution is quite significant 
and it is important to use the comprehensive TMD evolution rather than a separate evolution of the transverse and longitudinal dependences respectively. 
See the overview talk by Daniel Boer\cite{Boer} at this conference.

At last, I want to mention the first version of TMD PDFlib has already created\cite{Hautmann:2014kza}. 

\vspace*{-7pt}
\section{Summary and Outlook}

 In summary, I just want to emphasize that three dimensional imaging of the nucleon is
 a hot and fast developing topic in last years. 
 Many progresses have been made and many questions are open. 
 Especially in view of the running and planned facilities such as the electron-ion colliders, 
 we expect even rapid development in next years. 
 
 I apologize for many aspects that I could not cover such as TMDs and Wigner function, model calculations, nuclear dependences, 
 and hyperon polarization. The readers are referred to many interesting talks at this conference.

I thank X.N. Wang, Y.K. Song, S.Y. Wei, K.B. Chen, J.H. Gao and many other people for collaboration and help in preparing this talk.  
My sincere thanks also go to John Collins for communications.
This work was supported in part by NSFC 
(Nos.11035003 and 11375104),  and the Major State Basic Research Development Program in China (No. 2014CB845406).

\vspace*{-7pt}


\begin{thebibliography}{0}    


\bibitem{Fey72} R.P. Feynman, {\it Photon Hadron Interactions},  W.A. Benjamin. 1972.



\bibitem{Ellis:1982wd} 
  R.~K.~Ellis, W.~Furmanski and R.~Petronzio,
  Nucl.\ Phys.\ B {\bf 207}, 1 (1982).
%
%
\bibitem{Ellis:1982cd} 
 R.~K.~Ellis, W.~Furmanski and R.~Petronzio,
  Nucl.\ Phys.\ B {\bf 212}, 29 (1983).
  
\bibitem{Qiu:1990xxa} 
  J.~-w.~Qiu and G.~F.~Sterman,
  Nucl.\ Phys.\ B {\bf 353}, 105 (1991); 
%
  B {\bf 353}, 137 (1991).
 
 
\bibitem{Collins:1989gx} 
  J.~C.~Collins~{\it et al.,}   
  Adv.\ Ser.\ Direct.\ High Energy Phys.\  {\bf 5}, 1 (1988).
 



\bibitem{Sivers:1989cc}
  D.~W.~Sivers,
  Phys.\ Rev.\ D {\bf 41}, 83 (1990); {\bf 43}, 261 (1991).


\bibitem{Boros:1993ps} 
  C.~Boros, Z.~T.~Liang and T.~C.~Meng,
  Phys.\ Rev.\ Lett.\  {\bf 70}, 1751 (1993).
  

\bibitem{Collins:1992kk} 
  J.~C.~Collins,
  Nucl.\ Phys.\ B {\bf 396}, 161 (1993).


\bibitem{Brodsky:2002cx} 
  S.~J.~Brodsky, D.~S.~Hwang and I.~Schmidt,
  Phys.\ Lett.\ B {\bf 530}, 99 (2002).

\bibitem{Collins:2002kn} 
  J.~C.~Collins,
  Phys.\ Lett.\ B {\bf 536}, 43 (2002).

\bibitem{Ji:2002aa} 
  X.~d.~Ji and F.~Yuan,
  Phys.\ Lett.\ B {\bf 543}, 66 (2002).

\bibitem{Belitsky:2002sm} 
  A.~V.~Belitsky, X.~Ji and F.~Yuan,
  Nucl.\ Phys.\ B {\bf 656}, 165 (2003).



\bibitem{Georgi:1977tv}
  H.~Georgi and H.~Politzer,
  Phys.\ Rev.\ Lett.\  {\bf 40}, 3 (1978).

\bibitem{Cahn:1978se}
  R.~N.~Cahn,
  Phys.\ Lett.\ B {\bf 78}, 269 (1978).



\bibitem{Goeke:2005hb} 
  K.~Goeke, A.~Metz and M.~Schlegel,
  Phys.\ Lett.\ B {\bf 618}, 90 (2005).

\bibitem{Mulders} P. Mulders, talk in this conference, and lectures in 17th Taiwan nuclear physics summer school,
Aug. 25-28, 2014. 








\bibitem{Gourdin:1973qx} 
  M.~Gourdin,
  Nucl.\ Phys.\ B {\bf 49}, 501 (1972).

\bibitem{Kotzinian:1994dv} 
  A.~Kotzinian,
  Nucl.\ Phys.\ B {\bf 441}, 234 (1995).

\bibitem{Diehl:2005pc} 
  M.~Diehl and S.~Sapeta,
  Eur.\ Phys.\ J.\ C {\bf 41}, 515 (2005).

\bibitem{Bacchetta:2006tn} 
  A.~Bacchetta {\it et al.,}  
  JHEP {\bf 0702}, 093 (2007).


\bibitem{Arnold:2008kf} 
  S.~Arnold, A.~Metz and M.~Schlegel,
  Phys.\ Rev.\ D {\bf 79}, 034005 (2009).
  
\bibitem{Pitonyak:2013dsu} 
  D.~Pitonyak, M.~Schlegel and A.~Metz,
  Phys.\ Rev.\ D {\bf 89}, no. 5, 054032 (2014).
  
    
\bibitem{Liang:2006wp} 
  Z.~-t.~Liang and X.~-N.~Wang,
  Phys.\ Rev.\ D {\bf 75}, 094002 (2007).


\bibitem{Song:2010pf} 
  Y.~-k.~Song, J.~-h.~Gao, Z.~-t.~Liang and X.~-N.~Wang,
  Phys.\ Rev.\ D {\bf 83}, 054010 (2011).

  
  
  
\bibitem{Song:2013sja} 
  Y.~-k.~Song, J.~-h.~Gao, Z.~-t.~Liang and X.~-N.~Wang,
  Phys.\ Rev.\ D {\bf 89}, 014005 (2014).


  
  
\bibitem{Wei:2013csa} 
  S.~-y.~Wei, Y.~-k.~Song and Z.~-t.~Liang,
  Phys.\ Rev.\ D {\bf 89}, 014024 (2014).
  
\bibitem{Wei:2014pma} 
  S.~Y.~Wei, K.~b.~Chen, Y.~k.~Song and Z.~t.~Liang,
  arXiv:1410.4314 [hep-ph].
  
  \bibitem{YKSong} Y.K. Song, talk given at this conference.
  \bibitem{SYWei} S.Y. Wei, talk given at this conference.
  
  
\bibitem{Mulders:1995dh}
  P.~J.~Mulders and R.~D.~Tangerman,
  Nucl.\ Phys.\ B {\bf 461}, 197 (1996).

\bibitem{Boer:1997mf} 
  D.~Boer, R.~Jakob and P.~J.~Mulders,
  Nucl.\ Phys.\ B {\bf 504}, 345 (1997).



\bibitem{Lu:2011th} 
  Z.~Lu and I.~Schmidt,
  Phys.\ Rev.\ D {\bf 84}, 114004 (2011).


\bibitem{Collins:1981uk} 
  J.~C.~Collins and D.~E.~Soper,
  Nucl.\ Phys.\ B {\bf 193}, 381 (1981)
  [Erratum-ibid.\ B {\bf 213}, 545 (1983)];
%
  Nucl.\ Phys.\ B {\bf 194}, 445 (1982).






\bibitem{Collins:1984kg} 
  J.~C.~Collins, D.~E.~Soper and G.~F.~Sterman,
  Nucl.\ Phys.\ B {\bf 250}, 199 (1985).

\bibitem{Collins:1985ue} 
  J.~C.~Collins, D.~E.~Soper and G.~F.~Sterman,
  Nucl.\ Phys.\ B {\bf 261}, 104 (1985).




\bibitem{Ji:2004hz} 
  X.~d.~Ji, J.~P.~Ma and F.~Yuan,
  Phys.\ Lett.\ B {\bf 610}, 247 (2005).

\bibitem{Idilbi:2004vb} 
  A.~Idilbi, X.~d.~Ji, J.~P.~Ma and F.~Yuan,
  Phys.\ Rev.\ D {\bf 70}, 074021 (2004).

\bibitem{Ji:2004xq} 
  X.~d.~Ji, J.~P.~Ma and F.~Yuan,
  Phys.\ Lett.\ B {\bf 597}, 299 (2004).

\bibitem{Ji:2004wu} 
  X.~d.~Ji, J.~p.~Ma and F.~Yuan,
  Phys.\ Rev.\ D {\bf 71}, 034005 (2005).






\bibitem{Henneman:2001ev} 
  A.~A.~Henneman, D.~Boer and P.~J.~Mulders,
  Nucl.\ Phys.\ B {\bf 620}, 331 (2002).


\bibitem{Zhou:2008mz} 
  J.~Zhou, F.~Yuan and Z.~T.~Liang,
  Phys.\ Rev.\ D {\bf 79}, 114022 (2009).

\bibitem{Kang:2011mr} 
  Z.~B.~Kang, B.~W.~Xiao and F.~Yuan,
  Phys.\ Rev.\ Lett.\  {\bf 107}, 152002 (2011).

\bibitem{Aybat:2011zv} 
  S.~M.~Aybat and T.~C.~Rogers,
  Phys.\ Rev.\ D {\bf 83}, 114042 (2011).

\bibitem{Aybat:2011ge} 
  S.~M.~Aybat {\it et al.,}  
  Phys.\ Rev.\ D {\bf 85}, 034043 (2012).

\bibitem{Anselmino:2012aa} 
  M.~Anselmino, M.~Boglione and S.~Melis,
  Phys.\ Rev.\ D {\bf 86}, 014028 (2012).

\bibitem{Sun:2013hua} 
  P.~Sun and F.~Yuan,
  Phys.\ Rev.\ D {\bf 88}, no. 11, 114012 (2013).

\bibitem{Echevarria:2014xaa} 
  M.~G.~Echevarria~{\it et al.,}  
  Phys.\ Rev.\ D {\bf 89}, 074013 (2014).

\bibitem{Aidala:2014hva} 
  C.~A.~Aidala~{\it et al.,}  
  Phys.\ Rev.\ D {\bf 89}, 094002 (2014).

\bibitem{Kang:2014zza} 
  Z.~B.~Kang, A.~Prokudin, P.~Sun and F.~Yuan,
  arXiv:1410.4877 [hep-ph].

\bibitem{Echevarria:2014rua} 
  M.~G.~Echevarria, A.~Idilbi and I.~Scimemi,
  Phys.\ Rev.\ D {\bf 90},  014003 (2014).

\bibitem{Boer} D. Boer, talk given at this conference.




\bibitem{Anselmino:2005nn} 
  M.~Anselmino~{\it et al.,}  
  Phys.\ Rev.\ D {\bf 71}, 074006 (2005).

\bibitem{Anselmino:2007fs} 
  M.~Anselmino~{\it et al.,}  
  Phys.\ Rev.\ D {\bf 75}, 054032 (2007).
  
\bibitem{Schweitzer:2010tt} 
  P.~Schweitzer, T.~Teckentrup and A.~Metz,
  Phys.\ Rev.\ D {\bf 81}, 094019 (2010).

\bibitem{Signori:2014kda} 
  A.~Signori, A.~Bacchetta, M.~Radici and G.~Schnell,
  JHEP {\bf 1311}, 194 (2013).

\bibitem{Anselmino:2013lza} 
  M.~Anselmino~{\it et al.,}  
  JHEP {\bf 1404}, 005 (2014).


\bibitem{Efremov:2004tp} 
  A.~V.~Efremov~{\it et al.,}  
  Phys.\ Lett.\ B {\bf 612}, 233 (2005).

\bibitem{Collins:2005ie} 
  J.~C.~Collins~{\it et al.,}  
  Phys.\ Rev.\ D {\bf 73}, 014021 (2006).

\bibitem{Arnold:2008ap} 
  S.~Arnold~{\it et al.,}  
  arXiv:0805.2137 [hep-ph].


\bibitem{Anselmino:2005nn} 
  M.~Anselmino~{\it et al.,}  
  Phys.\ Rev.\ D {\bf 71}, 074006 (2005).

\bibitem{Anselmino:2008sga} 
  M.~Anselmino~{\it et al.,}  
  Eur.\ Phys.\ J.\ A {\bf 39}, 89 (2009).

\bibitem{Vogelsang:2005cs} 
  W.~Vogelsang and F.~Yuan,
  Phys.\ Rev.\ D {\bf 72}, 054028 (2005).

\bibitem{Bacchetta:2011gx} 
  A.~Bacchetta and M.~Radici,
  Phys.\ Rev.\ Lett.\  {\bf 107}, 212001 (2011).



\bibitem{Anselmino:2013vqa} 
  M.~Anselmino~{\it et al.,}  
  Phys.\ Rev.\ D {\bf 87}, 094019 (2013).


\bibitem{Zhang:2008nu} 
  B.~Zhang, Z.~Lu, B.~Q.~Ma and I.~Schmidt,
  Phys.\ Rev.\ D {\bf 77}, 054011 (2008).
  
\bibitem{Zhang:2008ez} 
  B.~Zhang, Z.~Lu, B.~Q.~Ma and I.~Schmidt,
  Phys.\ Rev.\ D {\bf 78}, 034035 (2008).
  
\bibitem{Barone:2008tn} 
  V.~Barone, A.~Prokudin and B.~Q.~Ma,
  Phys.\ Rev.\ D {\bf 78}, 045022 (2008).

\bibitem{Barone:2009hw} 
  V.~Barone, S.~Melis and A.~Prokudin,
  Phys.\ Rev.\ D {\bf 81}, 114026 (2010).
  
\bibitem{Lu:2009ip} 
  Z.~Lu and I.~Schmidt,
  Phys.\ Rev.\ D {\bf 81}, 034023 (2010).


\bibitem{Zhu:2011ir} 
  J.~Zhu and B.~Q.~Ma,
  Phys.\ Rev.\ D {\bf 82}, 114022 (2010).





\bibitem{Hautmann:2014kza} 
  F.~Hautmann~{\it et al.,}  
  Eur.\ Phys.\ J.\ C {\bf 74}, no. 12, 3220 (2014).

\end{thebibliography}
\end{document}